\documentclass[aps,pra,showpacs,twocolumn]{revtex4}
\usepackage{amssymb}
\usepackage{amsmath}
\usepackage{graphicx}
\usepackage{epsfig}

\begin{document}

\title{Peierls distorted chain as a quantum data bus for quantum state
transfer}
\author{M.X. Huo}
\affiliation{Department of Physics, Nankai University, Tianjin
300071, China}
\author{Ying Li}
\affiliation{Department of Physics, Nankai University, Tianjin
300071, China}
\author{Z. Song}
\email{songtc@nankai.edu.cn} \affiliation{Department of Physics,
Nankai University, Tianjin 300071, China}
\author{C.P. Sun}
\email{suncp@itp.ac.cn} \homepage{http://www.itp.ac.cn/~suncp}
\affiliation{Institute of Theoretical Physics, Chinese Academy of
Sciences, Beijing, 100080,China} \affiliation{Department of
Physics, Nankai University, Tianjin 300071, China}

\begin{abstract}
We systematically study the transfer of quantum state of electron
spin as the flying qubit along a half-filled Peierls distorted
tight-binding chain described by the Su-Schrieffer-Heeger (SSH)
model, which behaves as a quantum data bus. This enables a novel
physical mechanism for quantum communication with always-on
interaction: the effective hopping of the spin carrier between
sites $A$ and $B$ connected to two sites in this SSH chain can be
induced by the quasi-excitations of the SSH model. As we prove, it
is the Peierls energy gap of the SSH quasi-excitations that plays
a crucial role to protect the robustness of the quantum state
transfer process. Moreover, our observation also indicates that
such a scheme can also be employed to explore the intrinsic
property of the quantum system.
\end{abstract}

\maketitle

\section{Introduction}

In many protocols of quantum information processing, it is crucial to
transmit the quantum state of qubits with high fidelity \cite{Chuang}. While
various schemes of quantum state transfer (QST) were proposed and
demonstrated experimentally \cite{q-commu} for the optical system even with
atom ensemble \cite{Lukin, Sun}, people have tried the best to implement
this task based on the solid-state systems, which are believed as the best
candidates for the scalable quantum computing \cite{solid,Bose}. It was
recognized that the quantum spin chain \cite{Christandl,Shi} and the Bloch
electron system \cite{YS} with the artificial nearest neighbor (NN)
couplings can be used as quantum data buses to transfer quantum information
perfectly. In Ref. \cite{Shi}, we discovered that the spectrum-parity
matching is responsible for the perfectness of most protocols of QST.

However, these solid-state based schemes for implementing quantum data bus
are too artificial with very specially designed NN couplings and only
single-particle cases were considered. Though it is believed that such an
engineered quantum system will be realized in the future experiments, at
least, this difficulty should be overcome in theoretical aspect. To this
end, we have tried to employ the higher-dimensional system or the
complicated quantum network, such as the spin ladder \cite{Ying}. Though the
spin ladder is not an ideal medium serving as a perfect data bus, it shows
that the long-distance QST is possible via the multi-exciton system due to
the existence of the spin gap, which is much larger than the coupling
strength connecting two separated qubits to the two ends of the ladder
respectively.

A medium can be a robust quantum data bus if the fidelity of transferring a
quantum state between two sites attached to it approaches unity at a finite
temperature. These consideration motivates us to use more natural energy
gapped materials to act as a quantum data bus. In this paper, we study the
function of quantum data bus for the conducting polymers (polyacetylene),
modeled as a Peierls distorted tight-binding chain (PDTBC) or the
Su-Schrieffer-Heeger (SSH) model\ \cite{SSH,Yu}.\ It is well known that,
taking the coupling between electron and photon into account, such a system
exhibits an instability against the lattice distortion, which induces an
energy gap for the spectrum in half-filled case as a result of dimerization.
Therefore, the SSH chain is a good candidate of quantum data bus for
fermion, which is just similar to the spin ladder for spin \cite{Ying}.
Furthermore, such a practical system allows us to investigate the case with
larger particle number by performing analytical analysis for the two distant
separated probe points $AB$ since the SSH model can be solved exactly.


\begin{figure}[tbp]
\includegraphics[bb=50 220 553 630, width=6 cm]{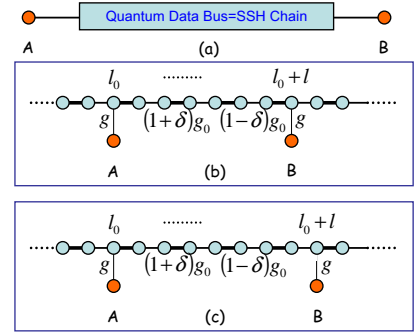}
\caption{(a) The schematic illustration of the system. Two sites A
and B connect to the two sites of PDTBC with the distance $l$
being (b) odd and (c) even. } \label{Model}
\end{figure}


As a protocol for the quantum communication with always-on interaction, two
points $A$ and $B$ are connected to two sites of the SSH chain (see the Fig.
1(a)). For the half-filled case, when the connections of $A$ and $\ B$\ with
SSH chain are switched off, the ground states are two-fold degenerate due to
the vanishing on-site chemical potential. When the connections switch on,
the two degenerate levels split while the eigenstates remain nearly
half-filled for the subsystem $AB$\ due to the energy gap of the SSH chain.
It leads to the effective interaction between $A$\ and $B$. Thus the level
spacing of two lowest eigenstates directly corresponds to the effective
hopping integral between $A$\ and $B$\ and determines the validity of
quantum state transfer via such a medium. Generally, if the energy gap is
large enough, an effective Hamiltonian $H_{AB}$\ of these two sites is
induced by this quantum data bus to perform perfect QST at low temperature.
Thus the validity of $H_{AB}$\ and the behavior of the effective hopping
integral (level spacing of two lowest eigenstates) as the distance between $%
A $ and $B$\ increases\ are crucial for the quality of a quantum data bus.
To demonstrate the properties of our scheme, we compare the eigenstates of $%
H_{AB}$\ with those reduced density matrices from the ground and the first
excited states of the total system.\textbf{\ }We find that the Peierls
energy gap of the SSH chain plays an important role in protecting the
robustness of the QST. The main conclusions are achieved both with the
analytical study and numerical calculation.

This paper is organized as follows. In Sec. II, the model setup of our
protocol and the spectrum of SSH chain are introduced. As a reasonable
approximation up to second order, the effective Hamiltonian $H_{AB}$ with
respect to the two separated points $A$ and $B$ is deduced by using the Fr\"{%
o}hlich transformation \cite{BCS, FRO} to \textquotedblleft
remove\textquotedblright\ the degree of freedom of SSH model in Sec. III. In
Sec. IV, the density matrices of qubits $A$ and $B$ for the ground and the
first excited states of the whole system are calculated to demonstrate the
quantum entanglement of them over a long distance. The QST scheme is also
investigated numerically in the region of the crossover between two types of
dimerization. It shows that, at this transition point, the QST becomes
fastest, but with the similar lost of fidelity to what has been discussed in
the Ref. \cite{Plenio}. In Sec.V, we\ discuss the quantum decoherence
problem due to the quasi-excitations of the SSH chain at finite temperature.
The summary and remarks are presented finally in Sec. VI.

\section{Model setup of quantum data bus based on the SSH chain}

In a polyacetylene, $\sigma $-electrons, which are localized between the two
bonded nuclei, connect one $C$ atom with the neighboring two $C$ and one $H$
atoms to form an elementary linear configuration; $\pi $-electrons, which
are less localized, behave as Bloch electrons. Such a system can be treated
by using tight-binding approximation and $\pi $-electrons can be the
candidate as a carrier of quantum information \cite{YS}. Here, we also
consider the coupling between electrons and dispersionless phonons
represented by local oscillators.

Let $u_{n}$ be the displacement of the $n$th $CH$ (the unit of a
polyacetylene) from the equilibrium position. Then, the tight-binding
Hamiltonian reads as
\begin{equation}
H_{SSH}=-\sum_{n,\sigma }g_{n+1}\left( c_{n+1,\sigma }^{\dagger }c_{n,\sigma
}+h.c.\right) +H_{C},  \label{Hssh}
\end{equation}%
where
\begin{equation}
H_{C}=\sum_{n}\left[ \frac{\Pi _{n}^{2}}{2M}+\frac{1}{2}K\left(
u_{n+1}-u_{n}\right) ^{2}\right]
\end{equation}%
describes the lattice motion. Here, $g_{n+1}=g_{0}-\lambda \left(
u_{n+1}-u_{n}\right) $ is the hopping integral and $\lambda $ is the
electron-lattice coupling constant. $\Pi _{n}$ denotes the canonical
momentum operator conjugate to $u_{n}$ and $M$ is the effective mass of the
unit of a polyacetylene (CH). $K$ is the spring constant due to $\sigma $%
-electrons, and $c_{j,\sigma }$ ($c_{j,\sigma }^{\dagger }$) is annihilation
(creation) operator of $\pi $-electrons on site $j$ with spin $\sigma =\pm 1$%
. This Hamiltonian is proposed by Su, Schrieffer and Heeger \cite{SSH} to
describe the dimerization phenomenon of the polyacetylene in association
with its conductivity.

In our scheme, we consider the QST between sites $A$ and $B$ connected to
the two sites of the SSH chain. Our purpose is to transfer a qubit state,
which is in a superposition state of electron spin up and spin down, through
the hopping of an polarized electron from the site $A$ to $B$ (see the Fig.
1(b)). Since the Hamiltonian (\ref{Hssh}) does not\ contain spin dependent
interaction, the spin polarization of every electron is conservative. In
this sense, the spatial motion of electrons along the chain will carry a
spin state from one location to another. Thus, we can study the fidelity of
spin state transfer by considering the fidelity of charge state transfer.
The connection of sites $A$ and $B$ to the SSH chain is described by the\
Hamiltonian
\begin{equation}
H_{I}=-g\sum_{\sigma }\left( c_{A,\sigma }^{\dagger }c_{l_{0},\sigma
}+c_{B,\sigma }^{\dagger }c_{l_{0}+l,\sigma }+h.c.\right) \text{,}
\end{equation}%
where $c_{A,\sigma }$ and $c_{B,\sigma }$ ($c_{A,\sigma }^{\dagger }$ and $%
c_{B,\sigma }^{\dagger }$) are annihilation (creation) operators of
electrons on sites $A$ and $B$ with spin $\sigma $, while $l_{0}$ and $%
l_{0}+l$ denote the connecting sites in the SSH chain.


\begin{figure}[tbp]
\includegraphics[bb=120 240 500 640, width=6 cm]{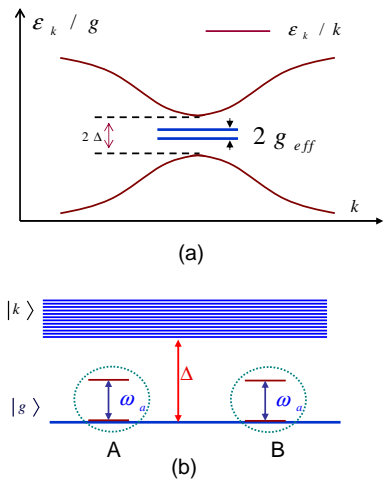}
\caption{(a) The single particle energy spectrum of dimerized polyacetylene.
(b) The eigenstates of the effective Hamiltonian about points $A$ and $B$.}
\label{SPS}
\end{figure}


In Fig. 2, we intuitively demonstrate how the above Hamiltonian can lead to
the effective hopping of electron between the additional sites $A$ and $B$.
We will also show how the effective hopping integral depends on the distance
between them. The original energy degeneracy at $E=0$ of the subsystem
formed by $A$ and $B$ will be removed by the switching on $g$ and then split
into two sub-levels with the level spacing $2g_{eff}$. Here, $g_{eff}$ is
the effective hopping integral, which can be obtained analytically as
follows.

To consider how a SSH chain can be used as a quantum data bus to transfer a
quantum state, we first summarize the known results about diagonalization of
the SSH Hamiltonian. The most important discovery is the prediction about
the Peierls transition happens when the polyacetylene is dimerized as $%
u_{n}=\left( -1\right) ^{n-1}u_{0}$. It minimizes the energy of the
one-dimensional electronic gas when the lattice vibrates slowly\ so that the
Born-Oppenheirmer approximation can determine an instantaneous eigen energy
of electron\ for a fixed lattice configuration. After dimerization, the
Hamiltonian then\ reads
\begin{equation}
H_{SSH}=-\sum_{n,\sigma }g_{0}\left[ 1-\left( -1\right) ^{n}\delta \right]
\left( c_{n+1,\sigma }^{\dagger }c_{n,\sigma }+h.c.\right) +E_{0}\text{,}
\end{equation}%
where $\delta =2\lambda u_{0}/g_{0}$ denotes the distortion of the hopping
integral\ and $E_{0}$ is a constant. Taking periodic boundary condition, it
can be diagonalized as
\begin{equation}
H_{SSH}=\sum_{k,\sigma }\epsilon _{k}\left( \alpha _{k,\sigma }^{\dagger
}\alpha _{k,\sigma }-\beta _{k,\sigma }^{\dagger }\beta _{k,\sigma }\right)
\text{,}
\end{equation}%
with the dispersion relation
\begin{equation}
\epsilon _{k}=2g_{0}\sqrt{\cos ^{2}\frac{k}{2}+\delta ^{2}\sin ^{2}\frac{k}{2%
}}  \label{dispersion}
\end{equation}%
for the excitations described by the fermion operators
\begin{equation}
\alpha _{k,\sigma }=\frac{1}{\sqrt{N}}\sum_{j=1}^{N/2}e^{-ikj}\left(
c_{2j-1,\sigma }-e^{i\theta _{k}}c_{2j,\sigma }\right)
\end{equation}%
and
\begin{equation}
\beta _{k,\sigma }=\frac{1}{\sqrt{N}}\sum_{j=1}^{N/2}e^{-ikj}\left(
c_{2j-1,\sigma }+e^{i\theta _{k}}c_{2j,\sigma }\right) \text{,}
\end{equation}%
where $N$ is the number of $CH$ in polyacetylene chain (or the length of the
chain), $k=4\pi m/N$, $m=0,1,2,\ldots ,N/2-1$, and
\begin{equation}
e^{i\theta _{k}}=\frac{g_{0}}{\epsilon _{k}}\left[ \left( 1+\delta \right)
+\left( 1-\delta \right) e^{-ik}\right] \text{.}
\end{equation}%
The single-particle spectrum (\ref{dispersion}) is illustrated in the Fig.
2(a) and the energy gap between the two bands is%
\begin{equation}
2\Delta =2\min \left\{ \epsilon _{k}\right\} =4g_{0}\delta \text{.}
\end{equation}

In next section, we will show that such an energy gap is necessarily
required for a desirable robust quantum data bus for QST. Usually, as
illustrated in Fig. 2(b), the large energy gap above the lowest two levels
can provide a kind of \textquotedblleft quantum protect\textquotedblright\
for the QST via the virtual excitations of the quantum data bus, which
induce a long range hopping of electron between sites $A$ and $B$ in the low
temperature. The dense continuity of spectrum above the gap ensures that the
strength of the effective hopping integral across $A$ and $B$ is so strong
that the fast entanglement can be generated. In the following, we will
demonstrate that such a dimerized gap plays a same role for QST as the spin
gap in spin ladder \cite{Ying}.

\section{Effective long range hopping induced by virtual quasi-excitations}

To deduce the effective Hamiltonian about the indirect coupling between two
attached sites $A$ and $B$, we utilize the Fr\"{o}hlich transformation,
whose original approach was used successfully for the BCS theory of
superconductivity. The effective Hamiltonian $H_{eff}=UHU^{-1}$ can be
achieved by a unitary operator $U=\exp \left( -S\right) $, where $H$ is the
Hamiltonian of the total system with the perturbation decomposition
\begin{equation}
H=H_{SSH}+H_{I}\text{.}  \label{H}
\end{equation}%
In the second order perturbation theory, we require the anti-Hermitian
operator $S$ obeys
\begin{equation}
H_{I}+\left[ H_{SSH},S\right] =0\text{.}  \label{S1}
\end{equation}%
Thus the effective Hamiltonian can be approximated as
\begin{equation}
H_{eff}\cong H_{SSH}+\frac{1}{2}\left[ H_{I},S\right] \text{.}
\end{equation}

There are two ways to connect the two sites $A$ and $B$\ to the SSH chain,
with (without) the mirror inversion symmetry as showed in Fig. 1(b) (Fig.
1(c)). Here, only the case with both $l_{0}$ and $l$ being odd is discussed.
The result can be used to all cases, because the system is invariant under
exchanging $A$ and $B$. As the solution of Eq. (\ref{S1}), $S$ can be
expressed explicitly as
\begin{eqnarray}
S &=&-\frac{g}{\sqrt{N}}\sum_{k,\sigma }\frac{1}{\epsilon _{k}}\left[ e^{ik%
\frac{l_{0}+1}{2}}c_{A,\sigma }^{\dagger }\left( \alpha _{k,\sigma }-\beta
_{k,\sigma }\right) \right. \\
&&-\left. e^{ik\frac{l_{0}+l}{2}-i\theta _{k}}c_{B,\sigma }^{\dagger }\left(
\alpha _{k,\sigma }+\beta _{k,\sigma }\right) -h.c.\right] ,  \notag
\end{eqnarray}%
which leads to the effective Hamiltonian
\begin{equation}
H_{eff}=H_{AB}+H_{0}
\end{equation}%
with
\begin{equation}
\lbrack H_{AB},H_{0}]=0.
\end{equation}%
Here
\begin{equation}
H_{AB}=\sum_{\sigma }g_{eff}\left( c_{A,\sigma }^{\dagger }c_{B,\sigma
}+c_{B,\sigma }^{\dagger }c_{A,\sigma }\right)  \label{HAB}
\end{equation}%
denotes the effective hopping of electron between $A$ and $B$ with the
strength
\begin{equation}
g_{eff}=\frac{2g^{2}}{N}\sum_{k}\frac{e^{-ik\frac{l-1}{2}+i\theta _{k}}}{%
\epsilon _{k}}\text{,}
\end{equation}%
while
\begin{eqnarray}
H_{0} &=&\sum_{k,\sigma }\epsilon _{k}\left( \alpha _{k,\sigma }^{\dagger
}\alpha _{k,\sigma }-\beta _{k,\sigma }^{\dagger }\beta _{k,\sigma }\right) +%
\frac{g^{2}}{2N}  \notag \\
&&\times \sum_{k,\sigma ,k^{\prime },\sigma ^{\prime }}\frac{1}{\epsilon
_{k^{\prime }}}\left[ A\left( k,k^{\prime }\right) (\alpha _{k,\sigma
}^{\dagger }\alpha _{k^{\prime },\sigma ^{\prime }}-\beta _{k,\sigma
}^{\dagger }\beta _{k^{\prime },\sigma ^{\prime }})\right.  \notag \\
&&+\left. B\left( k,k^{\prime }\right) (\beta _{k,\sigma }^{\dagger }\alpha
_{k^{\prime },\sigma ^{\prime }}-\alpha _{k,\sigma }^{\dagger }\beta
_{k^{\prime },\sigma ^{\prime }})+h.c.\right]
\end{eqnarray}%
describes the dynamics of the data bus, where
\begin{eqnarray}
A\left( k,k^{\prime }\right) &=&e^{-i\left( k-k^{\prime }\right) \frac{%
l_{0}+1}{2}}\left[ 1+e^{i\left( \theta _{k}-\theta _{k^{\prime }}\right) }%
\right] \text{,}  \notag \\
B\left( k,k^{\prime }\right) &=&e^{-i\left( k-k^{\prime }\right) \frac{%
l_{0}+1}{2}}\left[ 1-e^{i\left( \theta _{k}-\theta _{k^{\prime }}\right) }%
\right] \text{.}
\end{eqnarray}

Straightforward calculation shows that, in the thermodynamic limit, $%
N\longrightarrow \infty $, the hopping constant becomes
\begin{equation}
g_{eff}=\frac{g^{2}}{g_{0}}\frac{\left( -1\right) ^{\frac{l-1}{2}}}{1+\delta
}\left( \frac{1-\delta }{1+\delta }\right) ^{\frac{l-1}{2}}\text{.}
\label{teff-II}
\end{equation}%
Here, we have used the one-dimensional integral to replace the discrete sum
as $(2/N)\sum_{k}\longrightarrow \int_{0}^{2\pi }dk/2\pi .$

It needs to be pointed out that the effective Hamiltonian $H_{AB}$\ works
well only in the half-filled case. When the lower quasi-band is filled
fully, the next excitation should jump over the gap and then virtual
excitations can effectively induce the effective coupling. To demonstrate
the above result about the effective long range hopping induced by the SSH
chain, we schematically\ sketch the virtual process\ in Fig. 3, where the
quanta of data bus between the two sites are exchanged. If the energy gap of
the PDTBC is much larger than the connection couplings, $\Delta \gg g$, the
virtual transition is essentially second order process. There are two kinds
of virtual process for an initially half-filled state of the PDTBC. The
first one happens via the exchange of the electron in the upper quasi-band,
while the second one is based on the exchange of the hole in the lower
quasi-band. We would like to remind that the two quasi-bands will become
continuous in the thermodynamic limit. Nevertheless, there still exists
indirect interaction between $A$ and $B$ as the effective results of the
second order process.


\begin{figure}[tbp]
\includegraphics[bb=79 226 519 590, width=8 cm]{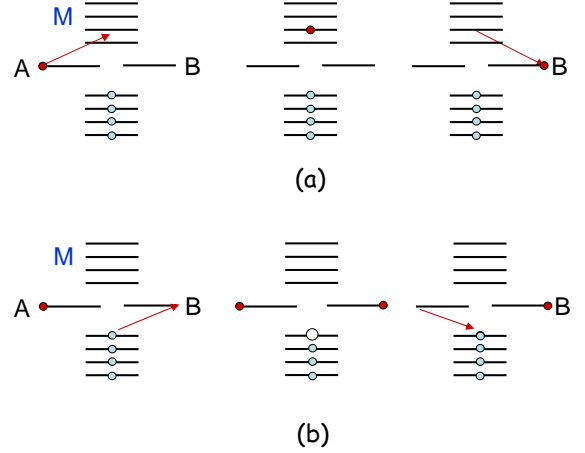}
\caption{Schematic illustration for the two processes of the second order
perturbation. The virtual transitions of (a) electrons between $A(B)$ and
the upper quasi-band and (b) holes between lower quasi-band and $A(B)$.
These processes lead to effective hopping of electron between $A$ and $B$
directly. }
\label{TOP}
\end{figure}


Now we consider the scheme using the SSH\ chain to transfer the quantum
state via the above effective virtual excitation process. Let Alice hold an
electron with spin state
\begin{equation}
\left\vert \varphi \right\rangle =\cos \frac{\theta }{2}\left\vert
\downarrow \right\rangle +e^{i\phi }\sin \frac{\theta }{2}\left\vert
\uparrow \right\rangle
\end{equation}%
at the site $A$, where $\left\vert \uparrow \right\rangle $ ($\left\vert
\downarrow \right\rangle $) denotes the spin up (down) state. Thus, the
initial state $\left\vert \Psi \left( 0\right) \right\rangle =\left\vert
\varphi \right\rangle _{A}\otimes \left\vert 0\right\rangle _{B}$ of the
total system is
\begin{equation}
\left\vert \Psi \left( 0\right) \right\rangle =\left( \cos \frac{\theta }{2}%
c_{A,\downarrow }^{\dagger }+e^{i\phi }\sin \frac{\theta }{2}c_{A,\uparrow
}^{\dagger }\right) \left\vert 0\right\rangle _{AB}\text{.}  \notag
\end{equation}%
Here, $\left\vert 0\right\rangle _{A}$ $(\left\vert 0\right\rangle _{AB})$
denotes the empty state, i.e., there is no electron at site $A$ (both sites $%
A$ and $B$).\ At the instant
\begin{equation}
t=\tau =\frac{\pi }{2\left\vert g_{eff}\right\vert }\text{,}
\end{equation}%
the total system evolves into a new factorized state $\left\vert \Psi \left(
\tau \right) \right\rangle =\left\vert 0\right\rangle _{A}\otimes \left\vert
\varphi \right\rangle _{B}$ or
\begin{equation}
\left\vert \Psi \left( \tau \right) \right\rangle =\left( \cos \frac{\theta
}{2}c_{B,\downarrow }^{\dagger }+e^{i\phi }\sin \frac{\theta }{2}%
c_{B,\uparrow }^{\dagger }\right) \left\vert 0\right\rangle _{AB}\text{,}
\notag
\end{equation}%
to realize a perfect quantum swapping. Then Bob at the site $B$ can receive
an electron with spin state $\left\vert \varphi \right\rangle $. Here,$\
\tau $ determines the characteristic time of quantum state transfer between
two locations $A$ and $B$, so the behavior of $\left\vert g_{eff}\right\vert
$ vs. $l$ is crucial for quantum information transfer.

It is desirable that the above scheme for QST based on the SSH virtual
excitation can work well over a longer distance, namely, the effective
hopping integral $g_{eff}$ should not decay too fast as the distance of two
points $AB$ increases. According to Eq. (\ref{teff-II}), $\left\vert
g_{eff}\right\vert $ decays exponentially as $l$ increases. However, $\sqrt{%
\left( 1-\delta \right) /\left( 1+\delta \right) }$ closes to $1$ for small $%
\delta $. So $\left\vert g_{eff}\right\vert $ does not decay rapidly for any
finite $l$. To demonstrate this, numerical simulation is performed. The
value of $\left\vert g_{eff}\right\vert $\ is computed from the energy
spacing between the ground and first excited states of the total system by
exact diagonalization method for the finite system with $N=500$, $l=10n+9$, $%
n\in \lbrack 0,7]$, $\delta =0.01$, $g=0.01$, and $g_{0}=1.0$. The exact
numerical result and the analytical result obtained from Eq. (\ref{teff-II})
are\ plotted in Fig. 4. It shows that they are in agreement with each other
well and $\left\vert g_{eff}\right\vert $ exhibits power law decay for
finite $l$. It indicates that the characteristic time $\tau $\ is
proportional to the distance $l$, which is crucial for the scalable quantum
information processing.


\begin{figure}[tbp]
\includegraphics[bb=36 211 529 679, width=7 cm]{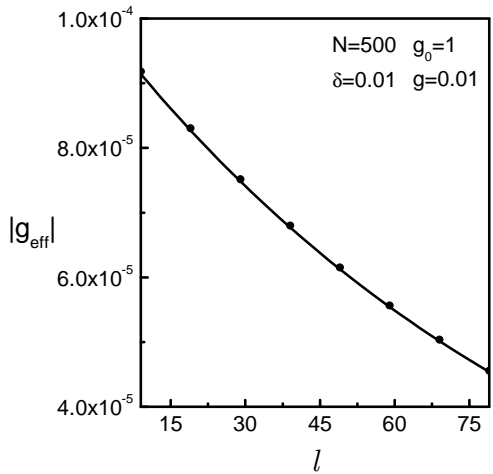}
\caption{The magnitude of effective hopping integral $g_{eff}$\
for the finite systems obtained by numerical exact diagonalization
(solid dot) and the approximate analytical results from Eq.
\protect\ref{teff-II} (line). Here only odd $l$ is plotted. It
shows that $g_{eff}$\ decays not so fast in this range of $l$ and
the exact numerical and approximate analytical results are in
agreement with each other very well.} \label{teff}
\end{figure}


\section{Reduced Density Matrix and Entanglement}

In the above section, we studied the behavior of $\left\vert
g_{eff}\right\vert $\ changing as the distance $l$\ increases. It determines
the speed of the QST via this SSH chain. However, all the conclusions
obtained should be based on the fact that the effective Hamiltonian $H_{AB}$
given by (\ref{HAB})\ is valid. In this section, we investigate the validity
of the effective Hamiltonian $H_{AB}$ by comparing the eigenstates of $%
H_{AB} $ with the density matrix reduced from the ground and first excited
states of the total system (\ref{H}). Notice that the Hamiltonian (\ref{H})
does not contain the spin-spin interaction term. Then we can regard the
system as a spinless fermion system and then apply the result obtained
feasibly for the spinless case to the original system. Therefore, in the
following discussion we will ignore the spin degree of freedom for
simplicity.

We define the states $\left\vert n_{-},n_{+}\right\rangle _{AB}$ by
\begin{eqnarray}
\left\vert 1,0\right\rangle _{AB} &=&\frac{1}{\sqrt{2}}\left( a_{A}^{\dagger
}-a_{B}^{\dagger }\right) \left\vert 0\right\rangle _{AB}, \\
\left\vert 0,1\right\rangle _{AB} &=&\frac{1}{\sqrt{2}}\left( a_{A}^{\dagger
}+a_{B}^{\dagger }\right) \left\vert 0\right\rangle _{AB}  \notag
\end{eqnarray}%
to denote the eigenstates of $H_{AB}$ in the half-filled subspace, where $%
a_{A}^{\dagger }$, $a_{B}^{\dagger }$ are spinless fermion operator at sites
$A$ and $B$. Here, $n_{-}$ $(n_{+})$ is the number of particles in the
anti-bonding (bonding) state. Then we calculate the \textquotedblleft
fidelity\textquotedblright\ of eigenstate\ defined by $P_{n_{-}n_{+}}=Tr(%
\rho _{R}\rho _{n_{-}n_{+}})$ or
\begin{equation}
P_{n_{-}n_{+}}=\sum_{\eta }|\left\langle n_{-},n_{+}\right. \left\vert
n_{-},n_{+},\eta \right\rangle _{AB}|^{2}.
\end{equation}%
Obviously, it can be described as the expectation value of the effective
density matrix $\rho _{n_{-}n_{+}}=\left\vert n_{-},n_{+}\right\rangle
_{AB}\left\langle n_{-},n_{+}\right\vert $\ in the exact eigenstates $%
\left\vert n_{-},n_{+}\right\rangle $ of the total system with the reduced
density matrix $\rho _{R}=Tr_{SSH}(\left\vert n_{-},n_{+}\right\rangle
\left\langle n_{-},n_{+}\right\vert )$. Here, $\left\vert n_{-},n_{+},\eta
\right\rangle =\left\vert n_{-},n_{+}\right\rangle _{AB}\left\vert \eta
\right\rangle _{SSH}$ denotes the eigenstates of (\ref{H}) with $g=0$, while
$\left\vert \eta \right\rangle _{SSH}$ denotes the groundstate ($\eta =0$)
and excited eigenstates ($\eta =1,2,...$) of the half-filled SSH chain.
State $\left\vert n_{-},n_{+}\right\rangle $ is the ground or first excited
states of the whole system (the Hamiltonian (\ref{H}) with $g\neq 0$), which
are spanned by the states possessing the same parity as the states $%
\left\vert n_{-},n_{+};0\right\rangle =\left\vert n_{-},n_{+}\right\rangle
_{AB}\otimes \left\vert 0\right\rangle _{SSH}$.

On the other hand, it is known that the eigenstates of $H_{AB}$ with unity
concurrence are maximally entangled states \cite{Wootters,Qian}. Thus by
calculating the mode concurrences of two sites $A$ and $B$ for the ground
and first excited states of the total system (\ref{H}), one can verify
whether the exact solutions can be well approximated by that from the
effective Hamiltonian. If \textquotedblleft fidelities\textquotedblright\ or
concurrences is near unity, it can be concluded that the effective
Hamiltonian $H_{AB}$ is indeed valid for the perfect QST.

In the following discussion, we try to explicitly express the exact
eigenstates $\left\vert n_{-},n_{+}\right\rangle $ of the total system with
the help of the Gellmann-Low theorem \cite{Many body}. In quantum field
theory, this theorem is used successfully to obtain the ground state of the
interaction field from the ground state of the free field. The exact
eigenstates can be expressed with the \textquotedblleft input
field\textquotedblright\ $\left\vert n_{-},n_{+};0\right\rangle $ as
\begin{equation}
\left\vert n_{-},n_{+}\right\rangle =\frac{U\left( 0,-T\right) \left\vert
n_{-},n_{+};0\right\rangle }{e^{-iE_{\pm 0}T}\left\langle n_{-},n_{+}\right.
\left\vert n_{-},n_{+};0\right\rangle }\text{,}
\end{equation}%
for $T\rightarrow \infty \left( 1+i\epsilon \right) $. Here,
\begin{equation}
U\left( t,t_{0}\right) =\mathcal{T}\exp \left\{ -i\int_{t_{0}}^{t}dt^{\prime
}H_{I}^{\prime }\left( t^{\prime }\right) \right\}
\end{equation}%
is the time-evolution operator in the interaction picture, where $\mathcal{T}
$ is the time-ordering symbol and $H_{I}^{\prime }\left( t^{\prime }\right)
=\exp (iH_{SSH}t^{\prime })H_{I}\exp (-iH_{SSH}t^{\prime })$. Similarly, we
have the \textquotedblleft output field\textquotedblright\
\begin{equation}
\left\langle n_{-},n_{+}\right\vert =\frac{\left\langle
n_{-},n_{+};0\right\vert U\left( T,0\right) }{e^{-iE_{\pm 0}T}\left\langle
n_{-},n_{+};0\right. \left\vert n_{-},n_{+}\right\rangle }\text{.}
\end{equation}%
With these formal expressions for the exact eigenstates of the total
system,\ we can calculate reduced density matrix with\ respect to the $AB$%
-subsystem. Then\ the expectation value of the projection operators of $%
\left\vert n_{-},n_{+}\right\rangle _{AB}$ in $\left\vert
n_{-},n_{+}\right\rangle $ can be calculated as
\begin{equation}
P_{n_{-}n_{+}}=\frac{\left\langle n_{-},n_{+};0\right\vert Q\left\vert
n_{-},n_{+};0\right\rangle }{e^{-iE_{\pm 0}\left( 2T\right) }\left\vert
\left\langle n_{-},n_{+};0\right. \left\vert n_{-},n_{+}\right\rangle
\right\vert ^{2}}\text{,}  \label{pp}
\end{equation}%
where%
\begin{equation}
Q=\mathcal{T}\left\{ Q_{n_{-}n_{+}}\exp \left[ -i\int_{-T}^{T}dt^{\prime
}H_{I}^{\prime }\left( t^{\prime }\right) \right] \right\} \text{,}
\end{equation}%
and
\begin{equation}
Q_{n_{-}n_{+}}=\left\vert n_{-},n_{+}\right\rangle _{AB}\left\langle
n_{-},n_{+}\right\vert \otimes \mathbf{1}\text{.}
\end{equation}

Notice that the above equation can not be calculated directly, because there
is not an explicit expression for $\left\vert n_{-},n_{+}\right\rangle $
with respect to the well known basis vectors $\left\vert
n_{-},n_{+};0\right\rangle $. To get rid of this difficulty, we use the
normalization condition $\left\langle n_{-},n_{+}\right. \left\vert
n_{-},n_{+}\right\rangle =1$ or
\begin{equation}
1=\frac{\left\langle n_{-},n_{+};0\right\vert U\left( T,-T\right) \left\vert
n_{-},n_{+};0\right\rangle }{e^{-iE_{\pm 0}\left( 2T\right) }\left\vert
\left\langle n_{-},n_{+};0\right. \left\vert n_{-},n_{+}\right\rangle
\right\vert ^{2}}\text{.}
\end{equation}%
Then the Eq. (\ref{pp}) is rewritten as
\begin{equation}
P_{n_{-}n_{+}}=\frac{\left\langle n_{-},n_{+};0\right\vert Q\left\vert
n_{-},n_{+};0\right\rangle }{\left\langle n_{-},n_{+};0\right\vert U\left(
T,-T\right) \left\vert n_{-},n_{+};0\right\rangle }\text{.}
\end{equation}

If the connection between sites $AB$ and SSH\ chain is switched off, i.e.,
the hopping integral in $H_{AB}$ vanishes, states $\left\vert
1,0;0\right\rangle $ and $\left\vert 0,1;0\right\rangle $ are the
eigenstates corresponding to vanishing eigen energies. Thus we can expand $%
P_{n_{-}n_{+}}$, which equals to unity when $H_{AB}$ is exact, as the series
of $g$. The Dyson expansion of$\ U\left( T,-T\right) $ also gives the
numerator and the denominator in the form of the series of $g$. Comparing
the terms of each order with respect to small coupling $g$, the zero order
of $P_{n_{-}n_{+}}$ ($P_{n_{-}n_{+}}^{\left( 0\right) }$) is unity, while
the first order of $P_{n_{-}n_{+}}$ ($P_{n_{-}n_{+}}^{\left( 1\right) }$)\
is zero as a result of the particle number conservation, and second order of
$P_{n_{-}n_{+}}$ ($P_{n_{-}n_{+}}^{\left( 2\right) }$) is\ non-zero and can
be expressed as
\begin{equation}
P_{n_{-}n_{+}}^{\left( 2\right) }=-\sum_{k}\frac{2g^{2}}{\epsilon _{k}^{2}N}%
\left[ 1+\left( n_{-}-n_{+}\right) \cos \left( k\frac{l-1}{2}-\theta
_{k}\right) \right] \text{.}  \label{p2}
\end{equation}%
It approximately describes the difference between the eigenstates of $H_{AB}$
and those reduced states from the exact eigenstates of the total system. In
order to clearly characterize our model by the correction $%
P_{n_{-}n_{+}}^{\left( 2\right) }$, we have the upper bound of $%
P_{n_{-}n_{+}}^{\left( 2\right) }$ directly from (\ref{p2}) as
\begin{equation}
\left\vert P_{n_{-}n_{+}}^{\left( 2\right) }\right\vert \leq \frac{4}{N}%
\sum_{k}\frac{g^{2}}{\epsilon _{k}^{2}}\text{.}
\end{equation}%
In the thermodynamic limit, this upper bound becomes
\begin{equation}
\left\vert P_{n_{-}n_{+}}^{\left( 2\right) }\right\vert \leq \frac{g^{2}}{%
2g_{0}^{2}\delta }\text{.}
\end{equation}%
This inequality provides us a necessary condition for the validity of the
effective Hamiltonian $H_{AB}$. Actually, the necessary condition is $%
g^{2}/(2g_{0}^{2}\delta )\ll 1$, which requires the validity condition
\begin{equation}
g/(\sqrt{2}g_{0})\ll \sqrt{\delta }.  \label{valid cond}
\end{equation}%
On the other hand, the sufficient condition is $g/g_{0}\ll 1$ for this
scheme. Then, if we take $\delta =g/g_{0}\ll 1$, it should satisfy the
sufficient and necessary conditions under which the effective Hamiltonian
can work well.

In this paper, we take $\delta =g/g_{0}=0.01$ for the numerical calculation
to demonstrate the validity of this scheme. Furthermore, the behavior of $%
P_{n_{-}n_{+}}$ around the point $\delta =0$, in which region the above
condition (\ref{valid cond})\ is violated, is also investigated by numerical
method. It shows that $P_{n_{-}n_{+}}$\ deviate from unity within this
region. We will discuss this problem in detail in the following.

Moreover, the mode concurrence between $A$ and $B$ for the state $\left\vert
n_{-},n_{+}\right\rangle $ can be gained from $P_{n_{-}n_{+}}$. It could
also be used to characterize the validity of the effective Hamiltonian $%
H_{AB}$. The mode concurrence for the state $\left\vert
n_{-},n_{+}\right\rangle $ is

\begin{equation}
C_{n_{-}n_{+}}=\max \left\{ 0,\left\langle X\right\rangle
_{n_{-}n_{+}}+2\left\vert \left\langle Y\right\rangle
_{n_{-}n_{+}}\right\vert -1\right\} \text{,}
\end{equation}%
where $X=n_{A}+n_{B}-2n_{A}n_{B}$, $Y=a_{A}^{\dagger }a_{B}+a_{B}^{\dagger
}a_{A}$, and $\left\langle O\right\rangle _{n_{-}n_{+}}=\left\langle
n_{-},n_{+}\right\vert O\left\vert n_{-},n_{+}\right\rangle $ denotes the
average value for an arbitrary operator $O$ (\cite{Qian2}). Since states $%
\left\vert n_{-},n_{+}\right\rangle $ and $\left\vert
n_{-},n_{+}\right\rangle _{AB}\otimes \left\vert 0\right\rangle $, possess
the same symmetry with respect to the points $A$ and $B$, we have
\begin{equation}
\left\langle X\right\rangle _{n_{-}n_{+}}+2|\left\langle Y\right\rangle
_{n_{-}n_{+}}|=\left\langle X-2\left( n_{-}-n_{+}\right) Y\right\rangle
_{n_{-}n_{+}}.
\end{equation}%
Similarly, we have
\begin{equation}
Q_{n_{-}n_{+}}=\frac{1}{2}\left[ X-2\left( n_{-}-n_{+}\right) Y\right] \text{%
{}}
\end{equation}%
Under the validity condition (\ref{valid cond}), $P_{n_{-}n_{+}}$ is near
unity and $2P_{n_{-}n_{+}}-1$ should be larger than zero. So we have mode
entanglement
\begin{equation}
C_{n_{-}n_{+}}=2P_{n_{-}n_{+}}-1\text{.}
\end{equation}%
Obviously, when the effective Hamiltonian $H_{AB}$\ is valid, the sites $A$
and $B$ are maximally entangled for the ground and first excited states of
the whole system.

\begin{table}[tbp]
\begin{center}
\begin{tabular}{ccccc}
\hline\hline
$\ \ \ l$ \ \  & \ \ $1-P_{10}$\ \ \  & \ \ \ \ \ $P_{10}^{\left( 2\right) }$
\ \  & \ $1-P_{01}$ \  & $\ \ \ P_{01}^{\left( 2\right) }$ \ \  \\
& $(\times 10^{-3})$ & $(\times 10^{-3})$ & $(\times 10^{-3})$ & $(\times
10^{-3})$ \\ \hline
9 & $4.48$ & $4.53\ $ & $5.39$ & $5.38\ $ \\
19 & $0.41$ & $0$.$41$ & $0.46$ & $0.47$ \\
29 & $4.66$ & $4.70$ & $3.69$ & $3.68$ \\
39 & $0.38$ & $0.38$ & $0.46$ & $0.47$ \\
49 & $4.60$ & $4.64$ & $4.30$ & $4.30$ \\
59 & $0.50$ & $0.50$ & $0.45$ & $0.46$ \\
69 & $4.45$ & $4.49$ & $5.80$ & $5.80$ \\
79 & $0.67$ & $0.67$ & $0.44$ & $0.44$ \\ \hline\hline
\end{tabular}%
\end{center}
\caption{$P_{10}$\ and $P_{01}$\ for the finite system with $N=500$, $l=10n+9$%
, $n\in \lbrack 0,7]$, $\protect\delta =0.01$, $g=0.01$, and $g_{0}=1.0$
obtained by numerical exact diagonalization and from the approximately
analytical expression Eq. (\protect\ref{p2}). They are in agreement with
each other very well, which shows that the effective Hamiltonian $H_{AB}$
can well describe the states of points $A$ and $B$ in the ground and first
excited states of the total system.}
\end{table}


From the Eq. (\ref{teff-II}) about effective hopping integral $g_{eff}$, one
can see that when $l=4n+1$ $(n=0,$ $1,$ $2,$ $...)$, $\left\vert
1,0\right\rangle $\ is the ground state of the total system, while $%
\left\vert 0,1\right\rangle $\ is the ground state for $l=4n+3$, due to the
sign of $g_{eff}$ being positive or negative. We denote the expectation
value $P_{n_{-}n_{+}}$ of the projection operator for the ground and first
excited states as $P_{g}$\ and $P_{e}$, respectively.

To demonstrate the validity of the effective Hamiltonian of the sites $AB$,
numerical simulation is performed for the finite system. We calculate $P_{g}$%
\ and $P_{e}$\ for the finite system with $N=500$, $l=10n+9$, $n\in \lbrack
0,7]$, $\delta =0.01$, $g=0.01$, and $g_{0}=1.0$ by numerical exact
diagonalization and from the approximately analytical expression Eq. (\ref%
{p2}). The results are listed in Table 1. It shows that they are in
agreement well and the single-particle (half-filled) eigenstates of the
effective Hamiltonian $H_{AB}$ can well describe the states of points $A$
and $B$ in the ground and first excited state of the total system. So $%
H_{AB} $\textbf{\ }is valid when temperature\textbf{\ }$T\ll 4g_{0}\delta
/k_{B}$\textbf{.}

In the above studies, it is found that the magnitude of distortion $\delta $%
\ is crucial for the quantum data bus to perform QST and the speed of QST is
sensitive to the dimerization. There are two types of dimerization
corresponding to $\pm \left\vert \delta \right\vert $. It is interesting to
investigate what happens near the transition point $\delta =0$. In this
region the analytical results obtained above is no longer valid due to the
vanishment of the gap. Numerical simulation should be employed to calculate $%
g_{eff}$, $P_{g}$\ and $P_{e}$, which characterize the property of QST. The
numerical results obtained by exact diagonalization for the systems with $%
N=100$, $200$, $l=23$, $33$, $43$ is plotted in Fig. 5. It shows that when
the systems approach to the critical point $\delta =0$, the gap $\Delta E$
between the ground and first excited states has a sharp peak, while $P_{g}$
tends to $0.25$, $P_{e}$ remains unitary approximately with only a very
slight jump around the transition point. It indicates that although QST
becomes fast around the critical point, the fidelity decreases rapidly. This
phenomenon is very similar to that discovered by Ref. \cite{Plenio}. This
fact also implies that the features of two probing sites $A$ and $B$, such
as the fidelity (mode concurrence) and the recurrent time $\pi /\Delta E$,
reveal\ the intrinsic nature of the quantum data bus, a quantum phase
transition-like behavior of the SSH chain.

\begin{figure}[tbp]
\includegraphics[bb=28 197 497 626, width=4 cm]{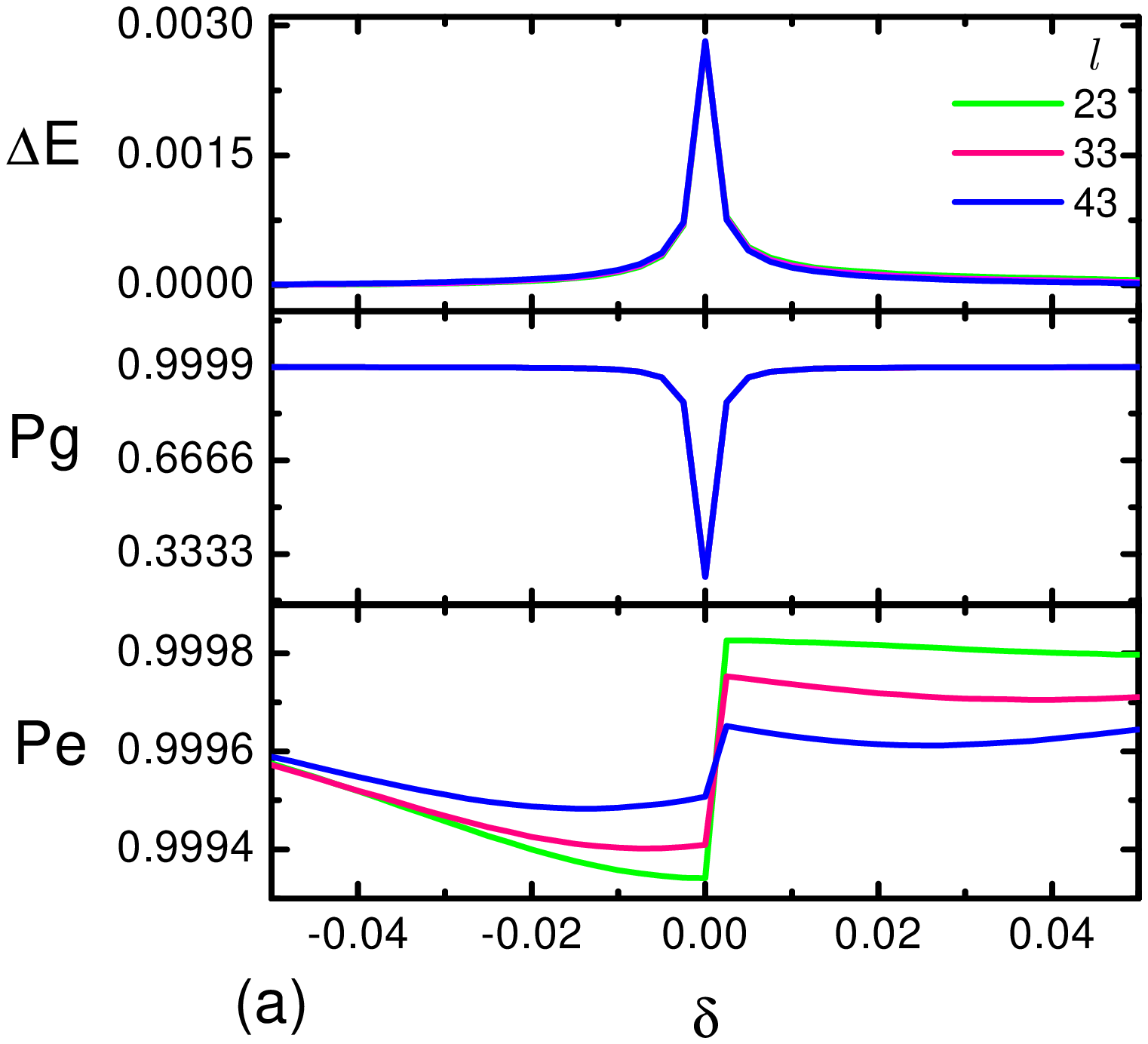} %
\includegraphics[bb=36 188 526 636, width=4 cm]{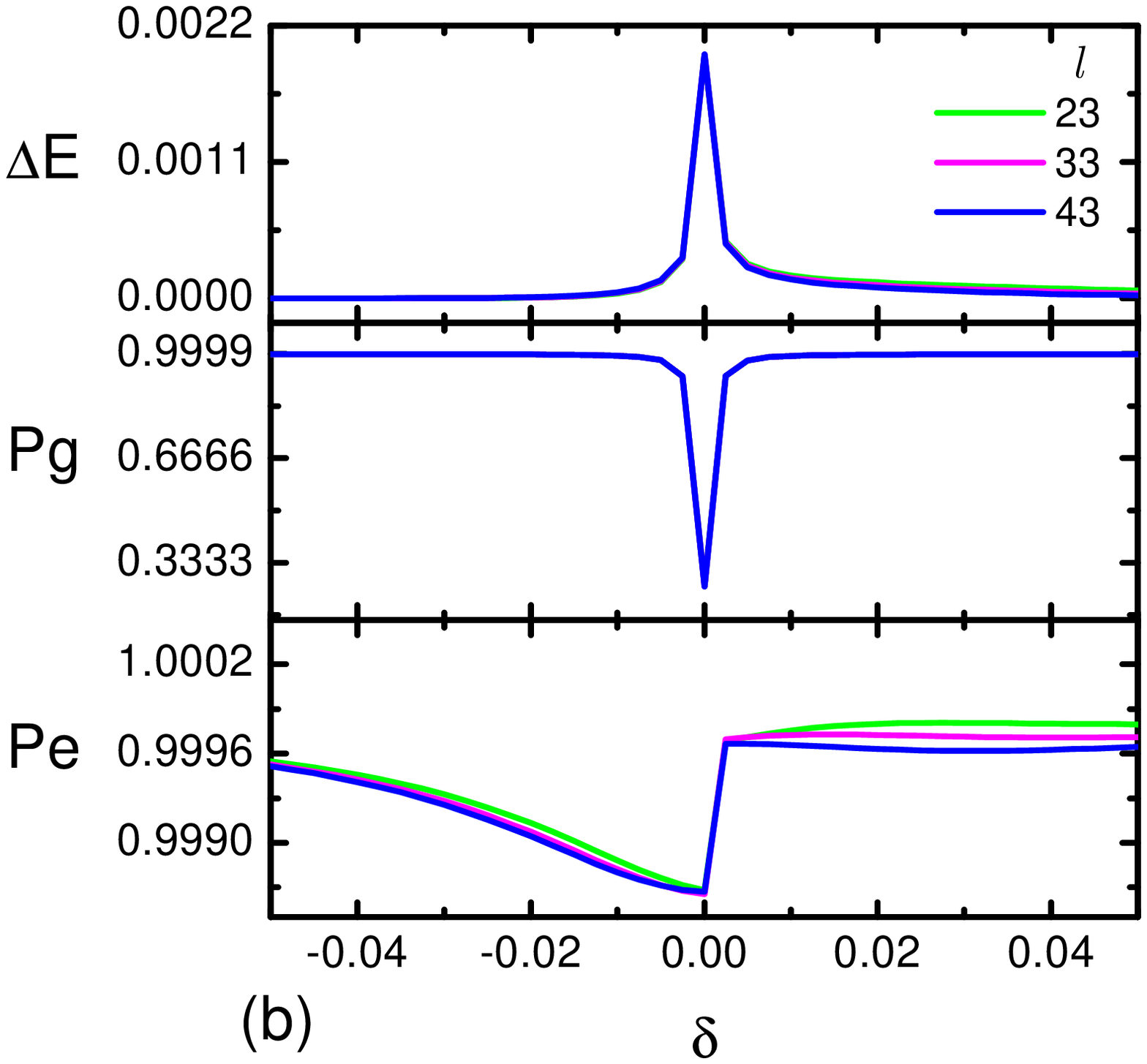}
\caption{The energy gap $\Delta E$ between the ground and first excited states, $%
Pg$, and $Pe$ of the system with (a) $N=100$; (b) $N=200$, and $l=23,33,43,$
obtained by numerical exact diagonalization as the function of $\protect%
\delta $. When $\protect\delta $ approaches to zero, the gap $\Delta E$ has
a sharp peak, while $P_{g}$ tends to $0.25$, $P_{e}$ remains unitary
approximately. It shows that although QST becomes fast, the fidelity
decreases rapidly around $\protect\delta =0$. This observation indicates
that the features of two sites $A$ and $B$ reveals\ the transition of the
data bus.}
\end{figure}

\section{Quantum decoherence by phonon excitations}

Finally we consider the decoherence problem due to the couplings with
auxiliary modes concerning phonon excitations. To this end, we need to
calculate the effect of electron-vibration coupling in the conducting
polymers, which is off-resonate to the electron motion. The
electron-vibration coupling gives rise to the disadvantage of the QST scheme
based on SSH\ chain, though it does not dissipate the energy of the
electronic subsystem.

Now we revisit the role of the electron-vibration coupling\ Hamiltonian (\ref%
{Hssh}) in the process of quantum information transfer based on the virtual
excitation of the SSH model. We notice that there is no direct interaction
between the electron motion and the quasi-excitation of SSH model when the
energy gap is much larger than the connection couplings since $%
[H_{AB},H_{0}]=0$. Thus the main source of decoherence is due to
electron-phonon couplings. This will realize a typical quantum decoherence
model for a two level system coupled to a bath of harmonic oscillators \cite%
{Sun,decohernce-Song}.

In order to analyze this decoherence problem more quantitatively, we
describe the single mode phonon by the perturbation
\begin{equation}
\delta =\gamma \left( b^{\dagger }+b\right) \text{,}
\end{equation}%
induced by the quantized\ vibration of the SSH\ chain,\ where $\gamma $ is a
constant and $b^{\dagger }$ ($b$) is the bosonic creation (annihilation)
operator of the phonon excitation. In\ small distortion case, according to (%
\ref{teff-II}) the effective Hamiltonian depends on the phonon excitation
through
\begin{equation}
g_{eff}\sim \omega _{s}\left[ 1-l\gamma \left( b^{\dagger }+b\right) \right]
\end{equation}%
where$\ $%
\begin{equation}
\omega _{s}=\left( -1\right) ^{\frac{l-1}{2}}\frac{g^{2}}{g_{0}}.
\end{equation}

Then we can rewrite the total Hamiltonian $H_{D}=H_{q}+H_{P}+H_{q-P}$
describing the quantum decoherence problem with
\begin{eqnarray}
H_{q} &=&\omega _{s}\sigma _{x}\text{,}  \notag \\
H_{P} &=&\omega _{0}b^{\dagger }b\text{,} \\
H_{q-P} &=&-l\omega _{s}\gamma \left( b^{\dagger }+b\right) \sigma _{x}\text{%
,}  \notag
\end{eqnarray}%
where the Hamiltonian $H_{q}+H_{q-P}$ corresponds to $H_{AB}$ and has been
rewritten in terms of the quasi-spin operators
\begin{eqnarray}
\sigma _{x} &=&a_{A}^{\dagger }a_{B}+a_{B}^{\dagger }a_{A}\text{,} \\
\sigma _{y} &=&-ia_{A}^{\dagger }a_{B}+ia_{B}^{\dagger }a_{A}\text{,}  \notag
\\
\sigma _{z} &=&a_{A}^{\dagger }a_{A}-a_{B}^{\dagger }a_{B}\text{.}  \notag
\end{eqnarray}%
Here, we have ignored the indices of the spin degree of freedom for avoiding
confusion, because all results we obtain in the following are unrelated to
spin.

Obviously, the effective electron-phonon coupling $H_{q-P}$ can only lead to
a phase shift in the spin qubit initially prepared in the quasi-spin states%
\begin{equation}
\left\vert \rightarrow \right\rangle =\left\vert 0,1\right\rangle _{AB}\text{%
, }\left\vert \leftarrow \right\rangle =\left\vert 1,0\right\rangle _{AB}%
\text{.}
\end{equation}%
It is also pointed out that the ground state (an eigenstate of the
quasi-spin $\sigma _{x}$) of the effective Hamiltonian\ is a mode
entanglement state. We consider the vibration mode initially prepared in a
thermal equilibrium state
\begin{equation}
\rho _{P}=\frac{1}{Z}\sum_{n}e^{-\frac{\omega _{0}n}{k_{B}T}}|n\rangle
\langle n|,
\end{equation}%
at the temperature $T$, where the partition function is%
\begin{equation}
Z=\frac{1}{1-e^{-\frac{\omega _{0}}{k_{B}T}}},
\end{equation}%
and $k_{B}$ is the Boltzman constant.

Let the electron be initially in a pure state $\left\vert \phi \right\rangle
=u\left\vert \rightarrow \right\rangle +v\left\vert \leftarrow \right\rangle
$. After a straightforward calculation we get the density matrix at time $t$
\begin{equation}
\rho \left( t\right) =U\left( t\right) (\left\vert \phi \right\rangle
\left\langle \phi \right\vert \otimes \rho _{P})U^{-1}\left( t\right)
\end{equation}%
and the corresponding reduced density matrix $\rho _{s}\left( t\right)
=Tr_{P}[\rho \left( t\right) ]$ of $AB$ can be obtained by tracing over the
phonon modes. The off-diagonal elements of $\rho _{s}\left( t\right) $ can
be given explicitly as
\begin{equation}
\rho _{s}^{\ast }\left( t\right) _{10}=\rho _{s}\left( t\right)
_{01}=uv^{\ast }D\left( T,t\right) \text{,}
\end{equation}%
where
\begin{equation}
D\left( T,t\right) =\exp \left[ -\alpha ^{2}\sin ^{2}\left( \frac{\omega
_{0}t}{2}\right) \right] \text{,}
\end{equation}%
is the so-called decoherence factor\cite{Sun,decohernce-Song}. The time
independent factor
\begin{equation}
\alpha ^{2}=2l^{2}\left( \frac{2g^{2}\gamma }{g_{0}\omega _{0}}\right)
^{2}\coth \left( \frac{\omega _{0}}{2k_{B}T}\right)
\end{equation}%
depends on the temperature $T$ and the distance $l$\ between $A$ and $B$.
The time dependent behavior of the decoherence factor is mainly determined
by the oscillating factors $\sin ^{2}\left( \omega _{0}t/2\right) $. It
indicates that, by increasing $l$ and $T$, $\alpha ^{2}$ becomes larger. It
leads to $D\left( T,t\right) \rightarrow 0$ except for some special instants
at
\begin{equation}
t=\frac{2m\pi }{\omega _{0}},m=0,1,2\ldots \text{.}
\end{equation}%
Here, $D\left( T,t\right) $ decays to a minimum $D_{\min }$ $=\exp \left(
-\alpha ^{2}\right) $ as temperature increases. This result shows that
thermal excitation of the phonons will block a perfect QST to some extent
and this effect becomes prominent as the distance between $A$\ and $B$\
increases\textbf{.} To consider the extent of this influence about the QST
scheme based on the SSH chain, we can calculate the fidelity or the purity $%
P=Tr_{P}[\rho _{s}\left( t\right) \rho _{0}\left( t\right) ]:$
\begin{equation}
P=1+2\left\vert uv^{\ast }\right\vert ^{2}\left[ D\left( T,t\right) -1\right]
\label{qqq}
\end{equation}%
with the ideal state%
\begin{equation}
\rho _{0}\left( t\right) =e^{-iH_{AB}t}\left\vert \phi \right\rangle \langle
\phi |e^{iH_{AB}t}.
\end{equation}%
The Eq. (\ref{qqq}) shows the intrinsic relation between the purity and the
decoherence factor.

Actually the off-diagonal elements are responsible for the intrinsic physics
of\ QST, because the time evolution driven by $H_{AB}$ for $A$ and $B$ will
stop if off-diagonal elements vanish. We notice that the norms of
off-diagonal elements is proportional to the decoherence factor $D\left(
T,t\right) $. In order to show the influence of the electron-vibration
coupling on the QST, 3D diagram of $D\left( T,t\right) $ is plotted in Fig.
6(a) for the case with $g_{0}=10^{4}\omega _{0}$, $g=10^{2}\omega _{0}$, $%
\gamma =10^{-2}$, $l=9$, $t\in \lbrack 0,7\pi /\omega _{0}]$, and $T\in
\lbrack k_{B}/4\omega _{0},10k_{B}/\omega _{0}]$. In Fig. 6(b) and (c), the
cross sections of the 3D diagram for $D\left( T,t\right) $ are plotted to
demonstrate the behavior of the decoherence factor. Actually, there exist
other modes of phonon excitations in the SSH\ chain. But in the case of
dimerization, the mode of phonon we described above dominates the quantum
decoherence and thus it is reasonable that we only focus on the dimerization
mode.


\begin{figure}[tbp]
\includegraphics[bb=65 274 490 637, width=7 cm]{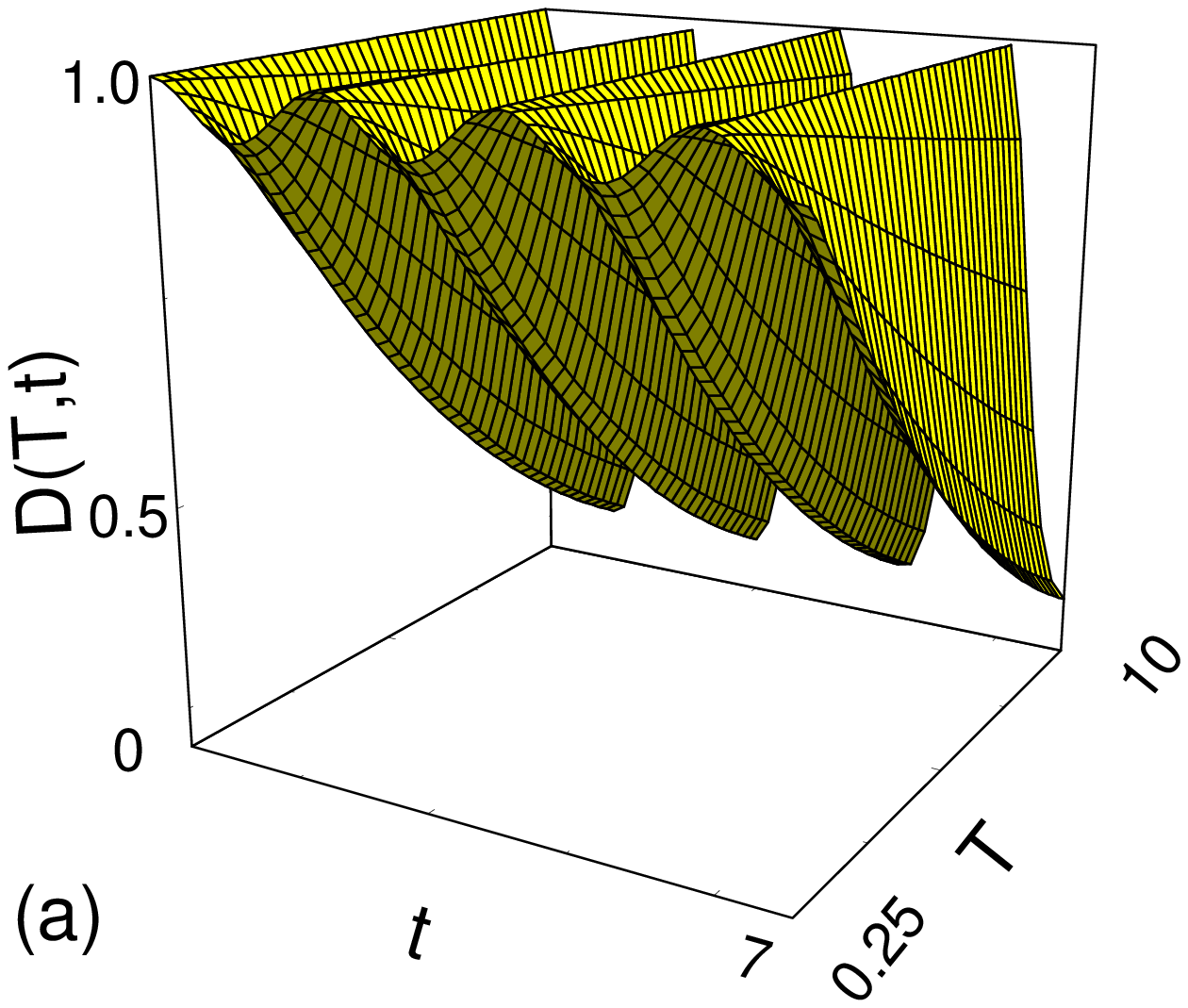} %
\includegraphics[bb=59 154 513 750, width=4 cm]{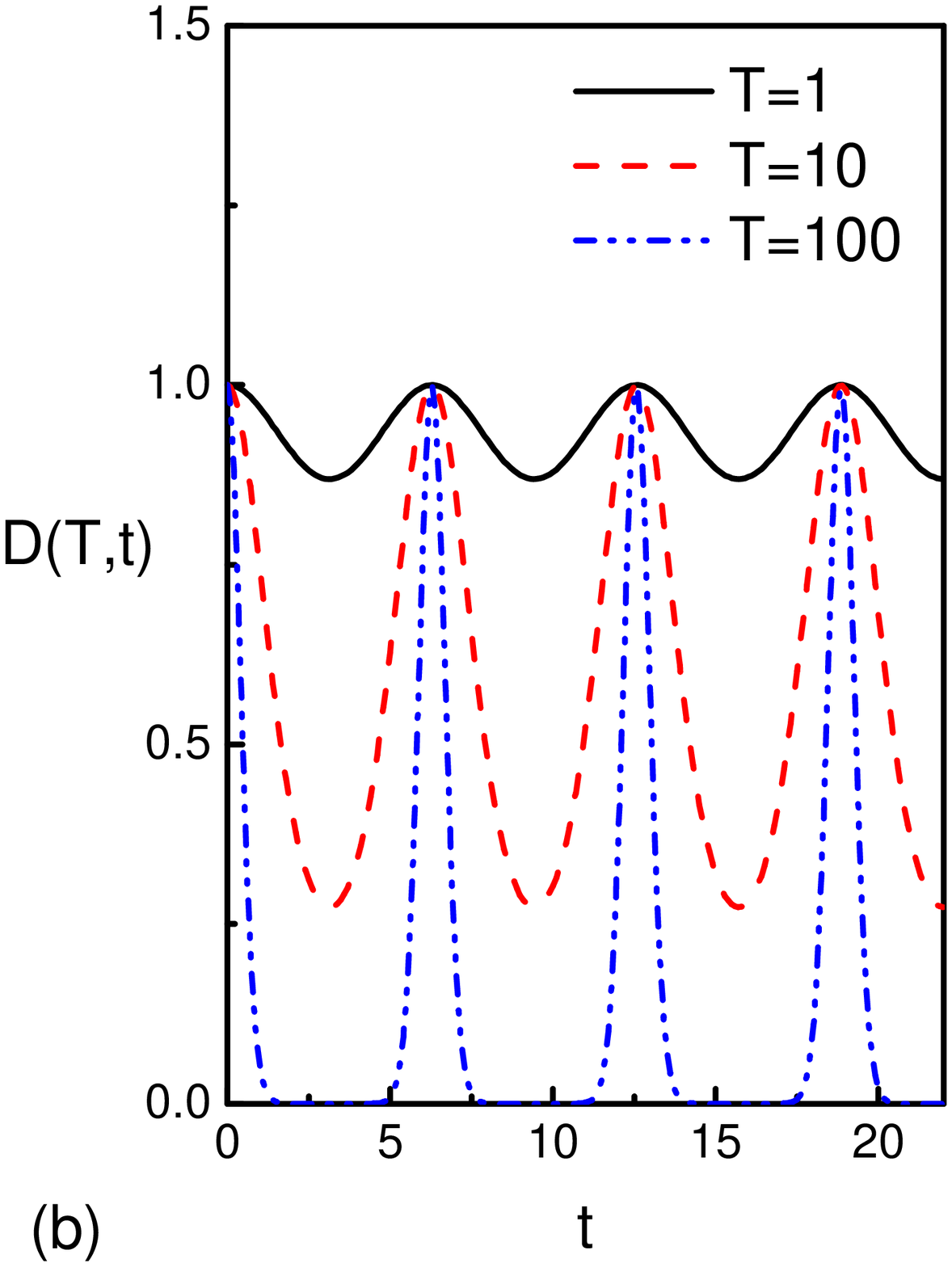} %
\includegraphics[bb=59 154 513 750, width=4 cm]{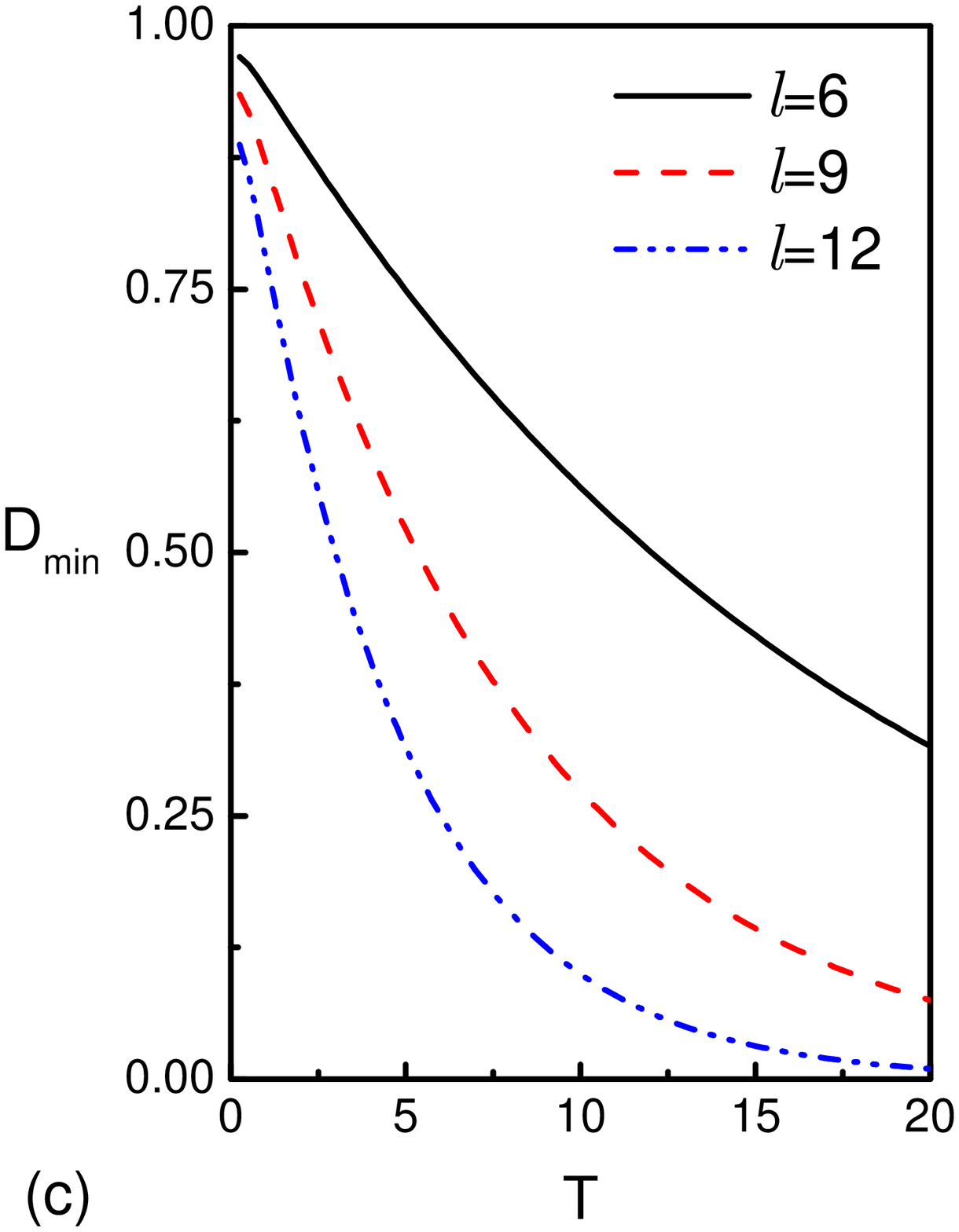}
\caption{(a) A 3D plot of the decoherence factor $D\left( T,t\right) $ as
the function of the temperature and time for the case with $g_{0}=10^{4}%
\protect\omega _{0}$, $g=10^{2}\protect\omega _{0}$, $\protect\gamma %
=10^{-2} $, $l=9$, $t\in \lbrack 0,7\protect\pi /\protect\omega _{0}]$, and $%
T\in \lbrack k_{B}/4\protect\omega _{0},10k_{B}/\protect\omega _{0}]$. (b),
(c) The cross section of $D\left( T,t\right) $ for different $T$ and $l$. It
shows that thermal excitation of the phonons will block a perfect QST as $T$
or $l$ increases.}
\end{figure}


\section{\textbf{Conclusions}}

In this paper we proposed a QST scheme based on a desirable quantum data
bus, which can be implemented by a more practical system, the conducting
polymers (polyacetylene) modeled as a SSH chain. This system serves as a
quantum data bus with the following advantages. Firstly, the strength of the
induced effective hopping of a qubit between two distant sites does not
decay rapidly over the distance between them, and thus it can realize a fast
quantum entanglement to transfer quantum information with the always-on
coupling and minimal control. Secondly, there exists an energy gap to ensure
the virtual excitation of the electrons in the SSH chain which induces the
effective hopping of a qubit between two distant sites, but does not
dissipate their energy. This shows the similar results to that of the spin
ladder \cite{Ying}. Thus a higher fidelity can be achieved for our QST
scheme based on the virtual excitation of the SSH chain in lower
temperature. Technologically, the more natural and less artificial designs
with practical modulation about coupling constants for our protocol push the
studies of QST based on real physical solid state system.

Furthermore, based on the analytical and numerical results, it is found that
the validity of $H_{AB}$\ strongly depends on the distortion $\delta $\ of
the SSH chain. Thus, as a measurable issue, it essentially reflects the
intrinsic property of the medium itself. In this sense, such a scheme can
also be employed to explore the intrinsic property of the quantum system.
These observations have some universality and may motivate us to investigate
the functions of quantum data bus based on other real physical systems.

We acknowledge the support of the NSFC (grant No. 90203018, 10474104,
10447133), the Knowledge Innovation Program (KIP) of Chinese Academy of
Sciences, the National Fundamental Research Program of China (No.
2001CB309310).

\end{document}